# Fast Sketch Segmentation and Labeling with Deep Learning

**Lei Li**
HKUST

**Hongbo Fu**
City University of Hong Kong

**Chiew-Lan Tai**
HKUST

We present a simple and efficient method based on deep learning to automatically decompose sketched objects into semantically valid parts. We train a deep neural network to transfer existing segmentations and labelings from 3D models to freehand sketches without requiring numerous well-annotated sketches as training data. The network takes the binary image of a sketched object as input and produces a corresponding segmentation map with per-pixel labelings as output. A subsequent post-process procedure with multi-label graph cuts further refines the segmentation and labeling result. We validate our proposed method on two sketch datasets. Experiments show that our method outperforms the state-of-the-art method in terms of segmentation and labeling accuracy and is significantly faster, enabling further integration in interactive drawing systems. We demonstrate the efficiency of our method in a sketch-based modeling application that automatically transforms input sketches into 3D models by part assembly.

Freehand sketching is frequently adopted as an efficient means of visual communication. Nowadays, the wide adoption of touch devices, together with the development of well-designed drawing software (e.g., Autodesk SketchBook), notably gives rise to easy creation of digital sketches without pen and paper. Unlike photos, which are faithful captures of the real world from cameras, sketches are artistic depictions from humans. Due to various levels of abstraction and distortion existing in sketches, the computer is still far from being able to robustly interpret their underlying semantics conveyed by humans.

Existing studies on sketch analysis, such as sketch classification or sketch-based retrieval, have mainly focused on interpreting an input sketch globally, lacking further understanding of its constituent parts. Sketch segmentation is a step towards finer-level sketch analysis.[1–3] Its goal is to decompose an input sketch into several semantically meaningful components, to which corresponding semantic labels may be assigned at the same time. Yet segmenting freehand sketches automatically is still a challenging task, because hand-crafted features or heuristic relations of the strokes designed for segmentation[1,2] may be sensitive to large variations of the sketches. Many existing sketch-based systems[4,5] require users to explicitly segment input drawings into meaningful components. An automatic and real-time sketch segmentation and labeling algorithm allows users to draw continuously without interruptions, paving the way to more natural human-computer interaction and downstream applications, such as sketch-based modeling by part assembly,[6] sketch editing[7] or sketch captioning.[8]

In this work, we focus on segmenting and labeling individually sketched objects. There has been research effort to investigate data-driven approaches for such a task,[1,2] more specifically, by transferring segmentations and labelings of 3D models[1] or 2D example sketches[2] to target sketched objects. However, these methods are either complicated or computationally inefficient even with small-scale databases serving as the knowledge base of semantic segmentation. The state-of-the-art method by Schneider and Tuytelaars[2] typically requires several minutes to interpret an input sketch. Thus, interactive sketching systems still cannot benefit from existing methods for more user-friendly interface designs.

We present a simple and efficient method based on deep Convolutional Neural Networks (CNNs) for semantic sketch segmentation and labeling. As illustrated in Figure 1, our network is trained to take the binary image of a sketched object as input and predict a corresponding segmentation map with per-pixel labelings as output. Our main challenge is the lack of a large volume of well-segmented freehand sketches with part annotations as training data. To address this, we utilize existing 3D model datasets with part segmentations and labelings.[1,9] We render each segmented 3D model from various viewpoints and extract edge maps to simulate human drawings. However, there exists a domain shift between edge maps from 3D models and freehand sketches from humans. Therefore, we adopt regularization techniques in our network design to improve the network performance on freehand sketches. Since sketches are commonly collected as sequences of polylines that can be viewed as graphs, we also perform a post-process with multi-label graph cuts[10] to further refine the segmentation result. Experiments show that our method is capable of effectively transferring the segmentation knowledge across the different domains. Our method outperforms the state-of-the-art method[2] in terms of segmentation accuracy and is approximately two orders of magnitude faster during test time.

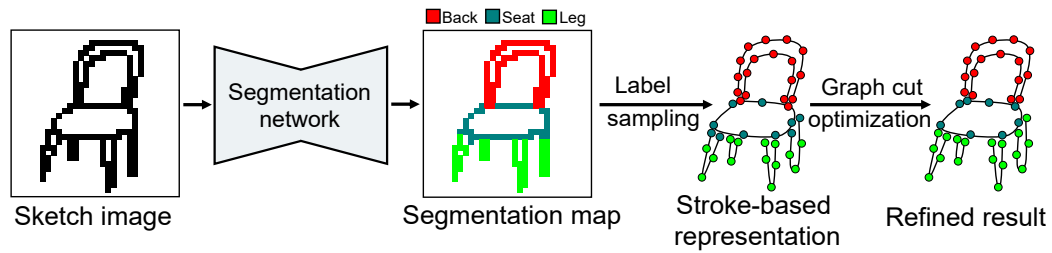

Figure 1. The pipeline of our method. The binary image of an input sketch is fed into our semantic sketch segmentation network to estimate a segmentation map of the constituent parts. Then we query part labels from the segmentation map for the stroke points in the stroke-based representation (sequences of polylines) of the input sketch and perform a post-process using multi-label graph cuts for further refinement.

We further demonstrate the application of our semantic segmentation method in a sketched-based modeling system, in which a completed sketch is automatically transformed into a 3D model by part assembly.[6] Specifically, once the user finishes drawing an object, our system automatically decomposes the sketch into semantic parts in a fraction of a second, retrieves similar 3D parts from a database of segmented 3D models and assembles them together. Thanks to the efficiency and accuracy of our semantic segmentation method, the user can instantly obtain 3D modeling results for further editing or refinement.

To sum up, our main contributions in this work are: 1) the first CNN-based approach for semantic segmentation and labeling of freehand sketches with better performance; 2) an application of the semantic segmentation method in sketch-based modeling by part assembly.

## RELATED WORK

*Sketch Segmentation and Labeling.* Early studies on sketch segmentation used low-level features of input drawings, such as distances, curvature or pen speed, to automatically decompose the inputs into geometric primitives or symbols.[11] By leveraging several low-level geometric features, Delaye and Lee[12] proposed an agglomerative clustering algorithm for online handwritten document segmentation, and Perteneder et al.[13] extended it to group sketches on large interactive screens, but semantic labelings are not considered. Noris et al.[7] developed Smart Scribbles, a user-guided segmentation system that combines the graph cut algorithm with constraints from additional annotations as strokes.

Recently a few studies have explored a data-driven approach to achieve high-level semantic segmentation of freehand sketches. To separate objects in a sketched scene, Sun et al.[3] employed a large clip-art database as the semantic knowledge base to merge strokes that belong to the same objects. However, their algorithm heavily depends on the drawing order of input strokes. To segment a single sketched object, Huang et al.[1] proposed to transfer part segmentations and labelings from a 3D model database by adopting a Mixed Integer Programming (MIP) formulation. However, their method needs manually specified viewpoints for input sketches for higher segmentation accuracy and requires nearly 40 minutes to process a single sketched object. A follow-up study by Schneider and Tuytelaars[2] used a Conditional Random Field (CRF) technique to transfer segmentations and labelings from a few example sketches to the inputs. It operates completely within the sketch domain and achieves high accuracy on the benchmark of Huang et al.[1] Yet their method still takes several minutes to segment a single sketch and may require a large number of manually segmented sketches as training data in practice to deal with large variations in freehand sketches, especially given the fact that an object can be drawn diversely under different viewpoints.

Our work is closely related to the studies by Huang et al.[1] and Schneider and Tuytelaars,[2] but our method can more efficiently predict sketch segmentations in a fraction of a second by running the inference pass of the trained network, instead of iterating over all the database models each time.[1] Besides, deep CNNs, adopted in our method for revealing part semantics of input sketches, do not require specially hand-crafted relations or features of input strokes.[1,2]

*Semantic Image Segmentation.* Studies on semantic image segmentation with deep learning are also related to our work. Their goal is to assign a label to each pixel of an input image of natural scenes. Long et al.[14] proposed to use fully convolutional networks for end-to-end learning, producing segmentation maps directly by one inference pass and thus yielding an efficient and unified framework. Several further improvements have been investigated as well, such as adding shortcut connections[15,16] or using dilated convolutions.[17] We adopt an encoder-decoder network design similar to the one by Ronneberger et al.[16] but transfer segmentations and labelings from edge maps of 3D models to 2D sketches, involving a domain shift.

Different from natural images with rich texture details, freehand sketches are highly abstract and only composed of simple strokes. Sarvadevabhatla et al.[8] designed a two-level CNN for parsing the image of a sketched object roughly as semantic regions, demonstrating the capability of neural networks in interpreting freehand sketches at part levels. However, as discussed in their work,[8] their region-based method cannot produce precise labelings of stroke pixels, that is, boundaries of the estimated part regions by their method may not correspond to the strokes of the input sketch.

## METHOD

Given a sketched object of a specific category in the stroke-based representation (i.e., sequences of polylines) as input, our method aims to decompose the sketch into semantically valid parts, to which corresponding labels are also assigned at the same time. We resort to deep CNNs, which are proven to have large capacity in learning descriptive features for various visual tasks given enough training data. Our designed network is trained to take a binary image of the sketched object as input and then build a hierarchical and global understanding of the input to produce a segmentation map with per-pixel labelings. We detail the network architecture in the following section. Training our semantic sketch segmentation network requires numerous well-segmented and labeled sketches, however existing large-scale crowd-sourced sketch datasets[18] lack such information. We use 3D models with segmentations (e.g., from ShapeNet[9]) to generate edge maps with part labelings. Our network is trained on the edge maps and then tested on freehand sketches, transferring segmentations and labelings across the two domains. We apply regularizations in the network design to avoid overfitting.

Sketches are often stored as sequences of polylines that can be directly gathered from the user's drawing trajectories. Given such a representation which can be treated as a graph, for each stroke point, we sample part labels from the segmentation map, estimated by our network, and perform multi-label graph cuts[10] to utilize the grouping information given by humans while drawing for further segmentation refinement.

## Network Architecture

We use an hourglass-shaped network that contains an encoder and a decoder for semantic sketch segmentation (see Figure 2).[16,19] The binary input image of the sketched object is of size 256×256, containing only one channel. The encoder passes the image through a sequence of convolutional layers, which perform progressive down-sampling to produce a relatively low-dimensional feature vector. The decoder inversely up-samples the output of the encoder via a series of up-convolutional layers to estimate a corresponding segmentation map as output.

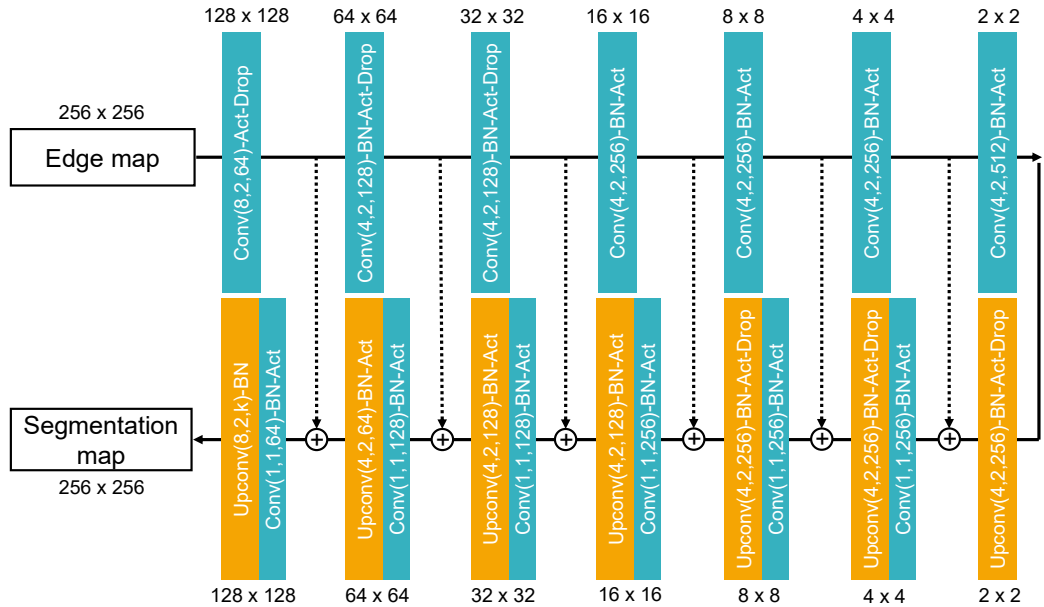

Figure 2. The architecture of our semantic sketch segmentation network. The upper part is an encoder that progressively down-samples the input while the lower part is a decoder that inversely up-samples feature representations. The input edge map is of size 256×256, so is the output segmentation map. The dashed lines represent shortcut connections and the symbol ⊕ denotes concatenation. (Conv: convolution; Act: activation; Drop: dropout; BN: batch normalization; Upconv: up-convolution.) The numbers within parentheses represent kernel size, stride and the number of output feature maps of a convolutional operation. The segmentation map contains *k* feature maps, representing estimations over *k* part labels (including background). During testing, the inputs are freehand sketches instead and the dropout operations are disabled.

More specifically, the encoder of our network contains seven convolutional layers with kernel size = 4 and stride = 2, except for the first layer with kernel size = 8 to accommodate the sparsity of stroke pixels. The number of output feature maps of each layer is shown in Figure 2. We apply batch normalization (except for the first layer) and leaky ReLUs with slope = 0.2 as activation functions after each convolutional operation. To better regularize the network and improve the robustness when the network deals with freehand sketches, we use dropout with probability = 0.5 in the first three layers during training. Note that features produced by the last layer are of size 2×2×512. We will use these features in the application section for sketch-based 3D model part retrieval.

Similarly, the decoder has seven up-convolutional layers, each with kernel size = 4 and stride = 2, except for the last layer with kernel size = 8. We apply batch normalization and ReLUs as activation functions (except for the last layer) after each up-convolutional operation. We use dropout with probability = 0.5 in the first three layers as well. To transfer information between corresponding network layers at the same level, we add shortcut connections, akin to the design of U-Net[12] for better information flow. Specifically, the input of each up-convolutional layer in the decoder is the concatenation of the outputs of its previous layer and the corresponding layer in the encoder. Additionally, before feeding the concatenation result into each up-convolutional layer, we pass it through a small module, which contains a stack of convolution (kernel size = 1, stride = 1), batch normalization and ReLU operations, to halve the number of feature maps. This helps to reduce the number of parameters of the decoder from 8.2M to 5.6M. The output segmentation map is of size 256×256 and contains $k$ channels, representing the estimations over $k$ part labels (including one label for blank background, i.e., non-stroke pixels). Note that the value of $k$ varies with object categories.

## Training

To train our network, we adopt the per-pixel cross-entropy loss function. Specifically, the softmax function is first applied to the $k$ channels at each pixel position of the predicted segmentation map. Let $\tilde{p}_i^j$ denote the probability estimation for the $j$ th part label ($1 \leq j \leq k$) at the $i$ th pixel position, and let $p_i^j$ be the one-hot representation of the ground truth (i.e., the bit corresponding to the ground truth label is 1 while the rest is 0). The loss function is defined in the following form:

$$L = \sum_i \sum_j -p_i^j \log(\tilde{p}_i^j) \quad .(1)$$

Here we briefly discuss alternative loss functions. During network design, we initially tried to introduce weighting factors in the loss function to balance the disproportional ratios between the background and the foreground as well as the ratios of part labels of an object category. However, we observed no significant improvements in segmentation accuracy. We also tried to consider only the segmentation predictions of the foreground, ignoring the background, in the loss function, but this modification did not improve the result either.

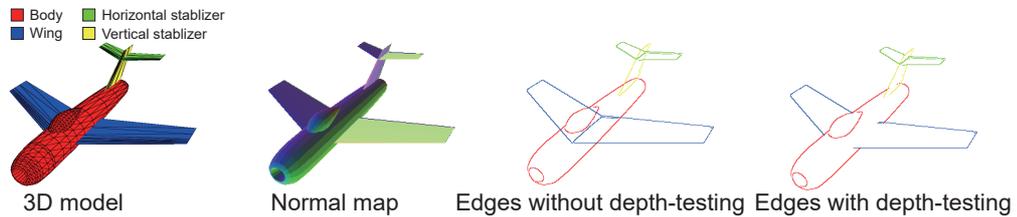

Figure 3. Example edge maps with ground-truth part segmentations and labelings derived from 3D models.

To generate training data, we render 3D models with part segmentations and labelings[1,9] of a specific object category to extract edge maps. The 3D models in the database are well aligned with consistent upright orientations. We sample viewpoints (36 ~ 72 for different object categories) on the upper unit viewing hemisphere, along with two camera-to-object distances (near and far), to render normal maps of the 3D models for Canny edge detection. Two types of edge maps are generated: with and without depth-testing (Figure 3), corresponding to the drawing styles of including or excluding hidden parts users may employ in freehand sketches. To obtain edge maps without depth-testing, we render the normal map and detect edges individually for each model part. We remove the invisible parts of the detected part edges by depth-testing to generate edge maps with depth-testing. Suggestive contours[20] are not used here, because the algorithm does not perform well on man-made models that are poorly triangulated.

We implement our semantic sketch segmentation network with Tensorflow. We use Adam ( $\beta_1 = 0.9$, $\beta_1 = 0.999$ ) for stochastic gradient descent update and set the learning rate to 0.0001. The network is trained for 80K steps with batch size = 32 on an NVIDIA GTX 1080Ti GPU. As suggested by Isola et al.,[19] during testing, we use batch normalization with the statistics of the testing data batch (freehand sketches) instead of the accumulated statistics of the training data batches (edge maps).

## Post-processing

Sketches are commonly collected as sequences of strokes, which can be viewed as initial segmentations. However, such grouping information introduced by humans is ignored in our segmentation network which processes the sketched object in a binary image format. Therefore, we perform a post-processing procedure exploiting the stroke-based representation to refine the network results (Figure 1). Specifically, we treat the sketch as a graph $G = (N, E)$, where the nodes $N$ are the stroke points and the edges $E$ connect sequentially adjacent points in a stroke. We define the following graph cut energy $L_G$ for optimization:

$$L_G = \sum_{p \in N} D_p(l_p) + \sum_{(p,q) \in E} V_{p,q}(l_p, l_q) \qquad .(2)$$

The first term is the data term, where $D_p(l_p)$ is the cost of assigning the point $p$ with part label $l_p$. We query the segmentation map estimated by the network and assign a constant cost $c_d$ if $l_p$ is not consistent with the label of the corresponding point of $p$ in the segmentation map, and a zero cost otherwise. The second term is the smoothness term, where $V_{p,q}(l_p, l_q)$ assigns a constant cost $c_s$ if $l_p$ and $l_q$ are different, and a zero cost otherwise. (Settings of $c_d$ and $c_s$ will be discussed in the evaluation section.) The energy minimization problem can be efficiently solved by the algorithm of Kolmogorov and Zabin.[10] This post-processing procedure helps to smooth out the noisy labelings produced by the network in each single stroke.

## EVALUATION

To evaluate our proposed semantic sketch segmentation method, we have performed experiments on two existing sketch datasets: Huang'14 dataset (10 object categories, 30 sketches per category via observational drawing by three participants)[1] and a subset of TU-Berlin dataset (250 object categories, 80 sketches per category via crowd sourcing).[18] Due to different collection methods, the sketches in the Huang'14 dataset closely resemble real-world objects with more complex structures while the sketches in the TU-Berlin dataset are more iconic and abstract.

Like existing studies,[1,2] we report segmentation accuracies using two types of evaluation metrics: pixel metric and component metric. For an input sketch, the pixel metric is calculated as the percentage of stroke pixels that are predicted with the same part labels as the ground truth. The component metric is calculated as the number of strokes with correctly predicted part labels over the total number of strokes in a sketch, irrespective of stroke length. A stroke is correctly labeled if the percentage of correctly labeled pixels in the stroke is above a certain threshold (75% in the experiments).[1,2] Figure 4 shows some sketch segmentation and labeling results produced by our method.

## Comparisons on Huang'14 Dataset

We compare our semantic sketch segmentation method with the MIP[1], CRF[2] and DeepLab[17] methods on the 10 object categories of Huang'14 dataset (Table 2 and Table 3). MIP[1] and CRF[2] are the traditional sketch segmentation methods, while DeepLab[17] is a state-of-the-art deep learning model for semantic image segmentation, which combines several techniques (e.g., atrous convolution and fully connected CRFs) compared to other deep learning models. The sketches in each category are already annotated with ground truth labelings.

To train our network on a specific category, we used the segmented 3D models provided by Huang et al.[1] to extract edge maps. However, the number of 3D models in each category is very limited (Table 1) and may not be enough for training large networks. Since other existing segmented 3D model datasets[9] contain incompatible part annotations with the ones provided by Huang et al.[1] (i.e., different segmentation granularity), we collected dozens of additional 3D models from 3D Warehouse and ShapeNet[9] for each category (Table 1) and manually segmented the new models with the same part labels used by Huang et al.[1] Additionally, we performed a simple data augmentation procedure by non-uniformly scaling the 3D models along each axis with factors 0.5 and 1.5 before rendering. We validated our network architecture on the chair sketches. For values of $c_d$ and $c_s$ used in the post-processing, we performed an exhaustive search on the validation data for the optimal setting, resulting in $c_d = 1$ and $c_s = 88$. After finalizing the network architecture and the parameters, we used the same design setting in the experiments for the other nine categories (and the experiments on the following TU-Berlin dataset). For completeness, we also include the performance of our method on the chair sketches in Table 2 and Table 3.

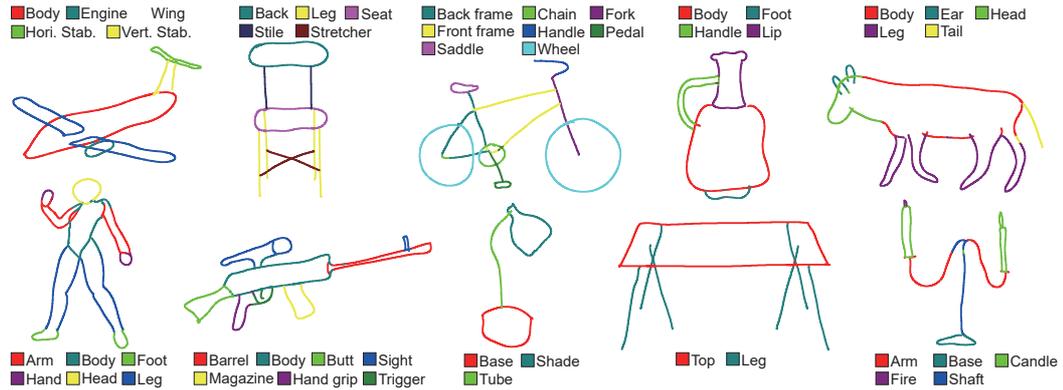

Figure 4. Automatically segmented and labeled sketches using our method.

Table 1. 3D models used for training our network.

| Category | For Huang'14 dataset | | | | | | | | | | For TU-Berlin dataset | | | | |
|---|---|---|---|---|---|---|---|---|---|---|---|---|---|---|---|
| | Airplane | Bicycle | Cdlbrm | Chair | Fourleg | Human | Lamp | Rifle | Table | Vase | Airplane | Chair | Guitar | Motorbike | Table |
| #Models | 59 | 37 | 35 | 68 | 20 | 17 | 20 | 25 | 57 | 61 | 500 | 500 | 500 | 202 | 500 |
| #Additional models | 16 | 26 | 32 | 33 | 43 | 27 | 25 | 28 | 24 | 28 | - | - | - | - | - |
| #Part lables | 6 | 9 | 6 | 11 | 5 | 6 | 3 | 7 | 7 | 4 | 4 | 4 | 3 | 6 | 2 |

Note: Cdlbrm stands for candelabrum.

Table 2 shows pixel metric accuracy comparisons of MIP-Auto,[1] CRF[2], DeepLab[17] and our method, which are fully automatic, and MIP-Manual,[1] which requires manually-specified viewpoints for input sketches. We simultaneously fed $x$ sketches ($x = 1, 2, 4, 6, 8, 10$, randomly divided) into our network as a batch for segmentation map prediction. For the last batch, we appended sketches that have already been tested to form a complete batch if necessary. We find that our method generally outperforms the CRF method with improvement of around 7-8% in pixel metric. Our method performs better than CRF in all the tested categories. Our method is approximately 12-13% higher than MIP-Auto, but requires additional 3D models as training data. However, MIP-Auto would not scale well with additional 3D models during testing, requiring even more running time due to its one-by-one iteration paradigm. Our method is even comparable to MIP-Manual that requires user assistance. Note that the CRF method used 20 sketches of a specific category as training data and the remaining 10 sketches as testing data, while the MIP methods were evaluated on all the sketches in each category. For the comparison of deep neural networks, the DeepLab[17] network, intended for semantic image segmentation, was trained with the same data (batch size = 4, 80K training steps) as ours. However, the experiment shows that it does not work well on the sketch input, because it focuses more on estimating regions of segmentation for the input image, struggling at region boundaries (i.e., thin edges like strokes).

Table 2. Pixel metric accuracy (%) on Huang'14 dataset.

| Category | Airplane | Bicycle | Cdlbrm | Chair | Fourleg | Human | Lamp | Rifle | Table | Vase | Average |
|---|---|---|---|---|---|---|---|---|---|---|---|
| MIP-Manual[1] | 82.4 | 78.2 | 72.7 | 76.5 | 80.2 | 79.1 | 92.1 | 75.9 | 79.1 | 71.9 | 78.8 |
| MIP-Auto[1] | 74.0 | 72.6 | 59.0 | 52.6 | 77.9 | 62.5 | 82.5 | 66.9 | 67.9 | 63.1 | 67.9 |
| CRF[2] | 55.1 | 79.7 | 72.0 | 66.5 | 81.5 | 69.7 | 82.9 | 67.8 | 74.5 | 83.3 | 73.2 |
| DeepLab[17] | 45.0 | 64.9 | 58.6 | 56.3 | 64.6 | 67.6 | 68.3 | 63.8 | 64.6 | 73.4 | 62.7 |
| Ours-NoGC | 81.1 | 82.9 | 74.7 | 70.0 | 79.6 | 74.8 | 85.7 | 68.5 | 77.6 | 81.1 | 77.6 |
| Ours-1 | 85.5 | 85.4 | 77.3 | 73.9 | 83.9 | 79.2 | 86.5 | 71.4 | 79.0 | **83.8** | 80.6 |
| Ours-2 | 85.2 | **86.3** | 77.0 | 72.2 | 84.2 | 79.8 | **88.2** | 71.5 | 79.2 | 82.6 | 80.6 |
| Ours-4 | **85.9** | 86.2 | 78.6 | 74.1 | 84.5 | 80.5 | 87.1 | 70.9 | 80.3 | 81.3 | 80.9 |
| Ours-6 | 85.8 | 86.0 | 78.5 | 75.0 | 84.8 | 80.3 | 87.0 | **71.9** | 80.6 | 82.1 | 81.2 |
| Ours-8 | 84.9 | **86.3** | **79.1** | 75.4 | 84.8 | **81.0** | 88.0 | 71.2 | 79.8 | 82.0 | 81.3 |
| Ours-10 | 85.7 | 86.2 | 78.3 | **75.5** | **84.9** | 80.5 | 87.8 | 70.6 | **81.4** | 82.7 | **81.4** |

Note: Ours-NoGC: our method without graph-cut optimization; Ours-$x$: $x$ sketches are simultaneously fed into our network as a batch. Best results among MIP-Auto, CRF, DeepLab and Ours in each column are in boldface.

Table 3. Component metric accuracy (%) on Huang'14 dataset.

| Category | Airplane | Bicycle | Cdlbrm | Chair | Fourleg | Human | Lamp | Rifle | Table | Vase | Average |
|---|---|---|---|---|---|---|---|---|---|---|---|
| MIP-Manual[1] | 66.2 | 66.4 | 56.7 | 63.1 | 67.2 | 64.0 | 89.3 | 62.2 | 69.0 | 63.1 | 66.7 |
| MIP-Auto[1] | 55.8 | 58.3 | 47.1 | 42.4 | 64.4 | 47.2 | 77.6 | 51.5 | 56.7 | 51.8 | 55.3 |
| CRF[2] | 48.7 | 68.6 | 66.2 | 61.6 | 74.2 | 63.1 | 77.2 | 65.1 | 65.6 | 79.1 | 67.0 |
| DeepLab[17] | 30.4 | 46.0 | 44.1 | 44.5 | 49.1 | 55.5 | 64.8 | 50.2 | 51.9 | 63.6 | 50.0 |
| Ours-NoGC | 65.4 | 67.9 | 59.2 | 60.5 | 66.5 | 61.9 | 78.1 | 56.3 | 67.3 | 71.9 | 65.5 |
| Ours-1 | 75.5 | 76.7 | 68.0 | 69.3 | **75.8** | 71.9 | 80.9 | 67.3 | 73.1 | **79.3** | 73.8 |
| Ours-2 | 76.2 | 76.8 | 68.2 | 68.1 | 75.2 | 72.3 | 83.1 | 67.6 | 74.7 | 77.1 | 73.9 |
| Ours-4 | 76.7 | 76.9 | 68.6 | 69.1 | 75.4 | 73.0 | 83.3 | 66.5 | 75.8 | 76.8 | 74.2 |
| Ours-6 | 76.7 | 76.9 | 69.4 | **70.3** | 75.1 | 72.9 | 82.9 | **67.9** | 76.4 | 77.0 | 74.6 |
| Ours-8 | 75.7 | 77.0 | **70.0** | 70.2 | 75.1 | **73.5** | 83.8 | 66.4 | 75.4 | 77.4 | 74.5 |
| Ours-10 | **76.9** | **77.1** | 69.9 | **70.3** | 75.5 | 72.8 | **83.8** | 65.7 | **77.3** | 78.0 | **74.7** |

Note: Best results among MIP-Auto, CRF, DeepLab and Ours in each column are in boldface.

Table 3 shows the component metric accuracies. Both MIP and CRF include a pre-processing procedure that splits each input stroke into segments at high-curvature or junction points and assume such segments are the basic units (i.e., as components) for assigning labels. As discussed by Schneider and Tuytelaars,[2] since the two methods may split strokes differently, the pixel metric comparison is more reliable than the component metric comparison. Our method does not require the pre-processing procedure, instead takes the whole sketch image as input to the network. Nevertheless, we used the sketches processed and split by Schneider and Tuytelaars[2] in the experiments for computing the component metric accuracies. Our method generally outperforms CRF with improvement of around 6-7% in component metric. Our method again consistently performs better than CRF in all the tested categories. The component metric also shows the inferiority of directly applying existing networks (like DeepLab[17]) of semantic image segmentation. From Table 2 and Table 3, it is observed that our method can obtain better results in both global (pixel metric) and fine-level (component metric) interpretation.

We evaluate the effect of the graph-cut post-processing procedure of our method by removing it in experiments (Ours-NoGC). Table 2 and Table 3 show that the accuracies drop both in pixel metric and component metric for Ours-NoGC. Additionally, we find that different batch sizes (Ours-$x$) during testing result in minor difference in the segmentation and labeling accuracies. This is due to the batch normalization operations whose statistics are computed only on the testing data batch (freehand sketches), without using the aggregated information from the training stage, to accommodate the domain shift. The experiments show that using larger batch sizes on the Huang'14 dataset gives slightly better results on average, though in practical applications (e.g., our sketch-based modeling), batch size is often set to one.

We performed the experiments using our Python implementation on a PC with an Intel Core i5 CPU @ 3.2GHz and 8 GB RAM. We quote the running time performance of the traditional MIP[1] and CRF[2] methods, both of which gave no detailed statistics for further comparison. The MIP method, implemented in MATLAB with the highly optimized library Gurobi, reported 40 minutes for interpreting a single sketch. The CRF method, which is the state-of-the-art and claims an order of magnitude faster than MIP, reported 2-3 minutes to test a sketch. It is implemented in MATLAB with the UGM library. We speculate their bottleneck might lie in the first stage involving dense SIFT feature extraction, GMM-based feature distribution estimation, and SVM-based classification for individual stroke segments. For fair comparison of running time, our implementation does not utilize GPU acceleration during testing. To process a single sketch, our method requires ~0.35s on average for segmentation map prediction and the post-process. In short, our method is around two orders of magnitude faster than the state-of-the-art method.

## Evaluation on TU-Berlin Dataset

To further evaluate the performance of our method, we also performed a similar set of experiments on a subset of the TU-Berlin dataset, which has more freely-drawn (80) sketch instances in each object category. We used a 3D model dataset provided by Yi et al.,[9] which has large-scale crowd-sourced segmentations and labelings, for edge map extraction. We selected five object categories, namely airplane, chair, guitar, motorbike, and table, by considering the category intersection of the two datasets, the number of part labels and the assembly-based modeling application. The numbers of 3D models and part labels in each category are listed in Table 1. Note that the crowd-sourced segmentation of 3D models provided by Yi et al.[9] generally contains fewer part labels than the one used by Huang et al.[1] Sketches of the same five categories are selected from the TU-Berlin dataset for testing. We manually annotated these sketches accordingly with part labels used in the 3D models.

Table 4 and Table 5 show the segmentation and labeling accuracies of our method on this test set using the pixel and component metrics. The full set of chair sketches in the TU-Berlin dataset is used as testing data. We find that our graph-cut post-processing procedure consistently improves the performance on the TU-Berlin dataset as well. However, batching more sketches as input to the network is not effective on this dataset. We speculate that this may be due to the difference in shape variations of sketches in Huang'14 and TU-Berlin datasets. The more freely drawn sketches in the TU-Berlin dataset deviate more from the training edge maps, which commonly contain regular lines, making the estimation of statistics in the batch normalization operations less reliable during testing.

Table 4. Pixel metric accuracy (%) on TU-Berlin dataset.

| Category | Airplane | Chair | Guitar | Motorbike | Table | Average |
|---|---|---|---|---|---|---|
| Ours-NoGC | 77.6 | 91.7 | 78.5 | 66.0 | 92.0 | 81.1 |
| Ours-1 | **82.1** | **95.7** | **81.4** | 70.8 | **94.0** | **84.8** |
| Ours-2 | 78.8 | 95.0 | 79.6 | 70.9 | 92.9 | 83.4 |
| Ours-4 | 78.2 | 94.8 | 79.8 | 71.1 | 92.7 | 83.3 |
| Ours-6 | 78.4 | 94.8 | 79.8 | 71.0 | 91.6 | 83.1 |
| Ours-8 | 79.1 | 94.5 | 79.6 | 71.3 | 92.6 | 83.4 |
| Ours-10 | 79.3 | 94.8 | 79.2 | **71.8** | 92.7 | 83.6 |

Note: Best results in each column are in boldface.

Table 5. Component metric accuracy (%) on TU-Berlin dataset.

| Category | Airplane | Chair | Guitar | Motorbike | Table | Average |
|---|---|---|---|---|---|---|
| Ours-NoGC | 63.5 | 89.5 | 67.0 | 47.5 | 87.3 | 71.0 |
| Ours-1 | **72.4** | **93.5** | **78.8** | 61.1 | **91.1** | **79.4** |
| Ours-2 | 70.8 | 92.9 | 77.2 | 61.5 | 90.5 | 78.6 |
| Ours-4 | 69.9 | 92.4 | 76.4 | 61.5 | 90.0 | 78.0 |
| Ours-6 | 70.3 | 92.8 | 76.7 | 61.8 | 89.4 | 78.2 |
| Ours-8 | 70.8 | 92.4 | 76.5 | 62.2 | 90.2 | 78.4 |
| Ours-10 | 70.7 | 92.7 | 75.9 | **62.6** | 90.3 | 78.4 |

Note: Best results in each column are in boldface.

## Limitations

Figure 5 shows some sketches with poor segmentation and labeling results by our method. Our method (mainly the network) may produce flipped segmentations due to the failure of interpreting correct viewpoints for the input sketches (Figure 5-(a)(b)). Certain levels of shape ambiguity in the inputs (Figure 5-(c), seat and back are of similar shapes) or sketches that are drastically different from 3D models in the training data (Figure 5-(d), no 3D human model with such a pose) may also pose challenges to our method.

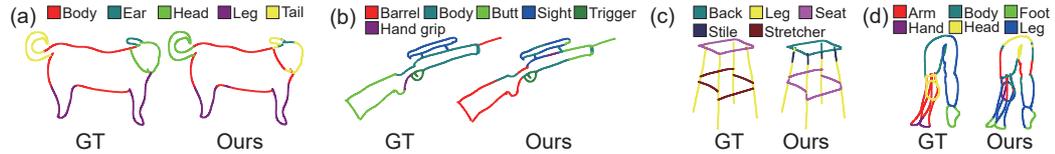

Figure 5. Sketches with poor segmentation and labeling results by our method. (GT is the ground truth segmentation.)

## APPLICATION

We demonstrate the efficiency of our semantic sketch segmentation method in a sketch-based modeling task that automatically converts input sketches into 3D models by assembling the recognized parts.[6] The efficiency of our method enables users to instantly obtain modeling results from freehand sketches and then perform further editing of the results.

Our sketch-based modeling task on a specific object category comprises an offline stage and an online stage. During the offline stage, for each category of 3D model parts, we extract feature vectors from the rendered edge maps of the model parts in the database (as described in the network architecture section). The extracted features are then stored for subsequent sketch-based part retrieval in the online stage. With the trained network for a specific object category, we use the features produced by the last layer of the encoder sub-network to quantify the input edge maps. Alternative features (e.g., bag-of-features) may also be employed. During the online stage, a complete user-sketched object is taken as input to our semantic segmentation method. Next, for each segmented part of the sketch, we again use the encoder sub-network to extract a corresponding feature vector, in the same way as above, for 3D model part retrieval. To search for similar 3D model parts within the identified part category, given the extracted feature vector of each sketch part, we compare the Euclidean distance between the feature vector and the ones extracted in the offline stage. The retrieval procedure finally produces a ranked list of candidate 3D model parts for each sketch part. We select the highest ranked from each list and assemble them together in a simple way just to demonstrate proof of concept as follows. We optimize the bounding boxes of these model parts such that their relative position and size relationships satisfy the configuration of 3D models from which the parts originated. The optimization is solved in a least-squares sense. More advanced assembly methods[6] can be used alternatively.

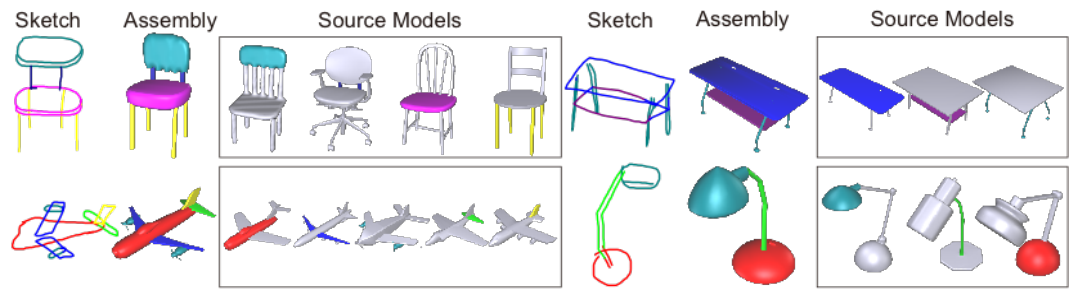

Figure 6. Example results of automatic sketch-based modeling by part assembly.

Figure 6 shows our sketch-based modeling results. Please refer to the supplementary video for modeling in action and more results. Other applications, such as sketch editing,[7] sketch captioning[8] or fine-grained sketch-based image retrieval, may also benefit from our fast semantic sketch segmentation method.

## CONCLUSION

In this work, we have proposed a fast semantic sketch segmentation method. We build upon deep neural networks to transfer segmentations and labelings from 3D models to freehand sketches, which is followed by a post-processing procedure with multi-label graph cuts for refinement. Experiments show that our proposed method outperforms the state-of-the-art method in terms of segmentation and labeling accuracy and is able to process a single sketch within a fraction of a second, which is around two orders of magnitude speedup. In the future, we will investigate methods that can further alleviate the domain shift problem between edge maps of 3D models and freehand sketches, such as incorporating adversarial training to improve the robustness of the segmentation network. Crowd-sourcing large-scale part annotations of freehand sketches may also ease the training process of the segmentation network.

## REFERENCES

1. Huang, Z., Fu, H. and Lau, R.W., 2014. Data-driven segmentation and labeling of freehand sketches. ACM Transactions on Graphics (TOG), 33(6), p.175.
2. Schneider, R.G. and Tuytelaars, T., 2016. Example-based sketch segmentation and labeling using crfs. ACM Transactions on Graphics (TOG), 35(5), p.151.

3. Sun, Z., Wang, C., Zhang, L. and Zhang, L., 2012, October. Free hand-drawn sketch segmentation. In European Conference on Computer Vision (pp. 626-639). Springer.
4. Xu, K., Chen, K., Fu, H., Sun, W.L. and Hu, S.M., 2013. Sketch2Scene: sketch-based co-retrieval and co-placement of 3D models. ACM Transactions on Graphics (TOG), 32(4), p.123.
5. Li, L., Huang, Z., Zou, C., Tai, C.L., Lau, R.W., Zhang, H., Tan, P. and Fu, H., 2016. Model-driven sketch reconstruction with structure-oriented retrieval. In SIGGRAPH ASIA 2016 Technical Briefs (p. 28). ACM.
6. Fan, L., Wang, R., Xu, L., Deng, J. and Liu, L., 2013, October. Modeling by drawing with shadow guidance. In Computer Graphics Forum (Vol. 32, No. 7, pp. 157-166).
7. Noris, G., Sýkora, D., Shamir, A., Coros, S., Whited, B., Simmons, M., Hornung, A., Gross, M. and Sumner, R., 2012. Smart scribbles for sketch segmentation. In Computer Graphics Forum (Vol. 31, No. 8, pp. 2516-2527).
8. Sarvadevabhatla, R.K., Dwivedi, I., Biswas, A. and Manocha, S., 2017. SketchParse: Towards Rich Descriptions for Poorly Drawn Sketches using Multi-Task Hierarchical Deep Networks. In Proceedings of the 2017 ACM on Multimedia Conference (pp. 10-18). ACM.
9. Yi, L., Kim, V.G., Ceylan, D., Shen, I., Yan, M., Su, H., Lu, C., Huang, Q., Sheffer, A. and Guibas, L., 2016. A scalable active framework for region annotation in 3d shape collections. ACM Transactions on Graphics (TOG), 35(6), p.210.
10. Kolmogorov, V. and Zabih, R., 2004. What energy functions can be minimizedvia graph cuts?. IEEE Transactions on Pattern Analysis & Machine Intelligence, (2), pp.147-159.
11. Gennari, L., Kara, L.B., Stahovich, T.F. and Shimada, K., 2005. Combining geometry and domain knowledge to interpret hand-drawn diagrams. Computers & Graphics, 29(4), pp.547-562.
12. Delaye, A. and Lee, K., 2015. A flexible framework for online document segmentation by pairwise stroke distance learning. Pattern Recognition, 48(4), pp.1197-1210.
13. Perteneder, F., Bresler, M., Grossauer, E.M., Leong, J. and Haller, M., 2015, November. cluster: Smart clustering of free-hand sketches on large interactive surfaces. In Proceedings of the 28th Annual ACM Symposium on User Interface Software & Technology (pp. 37-46). ACM.
14. Long, J., Shelhamer, E. and Darrell, T., 2015. Fully convolutional networks for semantic segmentation. In Proceedings of the IEEE conference on computer vision and pattern recognition (pp. 3431-3440).
15. Badrinarayanan, V., Kendall, A. and Cipolla, R., 2017. SegNet: A Deep Convolutional Encoder-Decoder Architecture for Image Segmentation. IEEE Transactions on Pattern Analysis & Machine Intelligence, (12), pp.2481-2495.
16. Ronneberger, O., Fischer, P. and Brox, T., 2015, October. U-net: Convolutional networks for biomedical image segmentation. In International Conference on Medical image computing and computer-assisted intervention (pp. 234-241). Springer.
17. Chen, L.C., Papandreou, G., Kokkinos, I., Murphy, K. and Yuille, A.L., 2018. Deeplab: Semantic image segmentation with deep convolutional nets, atrous convolution, and fully connected crfs. IEEE transactions on pattern analysis and machine intelligence, 40(4), pp.834-848.
18. Eitz, M., Hays, J. and Alexa, M., 2012. How do humans sketch objects?. ACM Trans. Graph., 31(4), pp.44-1.
19. Isola, P., Zhu, J.Y., Zhou, T. and Efros, A.A., 2017, July. Image-to-Image Translation with Conditional Adversarial Networks. In 2017 IEEE Conference on Computer Vision and Pattern Recognition (CVPR) (pp. 5967-5976). IEEE.
20. DeCarlo, D., Finkelstein, A., Rusinkiewicz, S. and Santella, A., 2003. Suggestive contours for conveying shape. ACM Transactions on Graphics (TOG), 22(3), pp.848-855.